\documentclass[aps,rpl,preprint,groupedaddress, showpacs]{revtex4}

\usepackage{graphicx}

\begin{document}

\title{Quantum heat engines and information}

\author{Ye Yeo}
\affiliation{Department of Physics, National University of Singapore, 10 Kent Ridge Crescent, Singapore 119260, Singapore}

\author{Chang Chi Kwong}
\affiliation{Department of Physics, National University of Singapore, 10 Kent Ridge Crescent, Singapore 119260, Singapore}

\begin{abstract}
Recently, Zhang {\em et al.} [PRA, {\bf 75}, 062102 (2007)] extended Kieu's interesting work on the quantum Otto engine [PRL, {\bf 93}, 140403 (2004)] by considering as working substance a bipartite quantum system $AB$ composed of subsystems $A$ and $B$.  In this paper,  we express the net work done $W_{AB}$ by such an engine explicitly in terms of the macroscopic bath temperatures and information theoretic quantities associated with the microscopic quantum states of the working substance.  This allows us to gain insights into the dependence of positive $W_{AB}$ on the quantum properties of the states.  We illustrate with a two-qubit XY chain as the working substance.  Inspired by the expression, we propose a plausible formula for the work derivable from the subsystems.  We show that there is a critical entanglement beyond which it is impossible to draw positive work locally from the individual subsystems while $W_{AB}$ is positive.  This could be another interesting manifestation of quantum nonlocality.
\end{abstract}


\maketitle

Heat engines are devices that extract energy from its environment in the form of heat and do useful work.  At the heart of every heat engine is a working substance, such as a gas-air mixture in an automobile engine.  The operation of the heat engine is achieved by subjecting the working substance of the engine to a sequence of thermodynamic processes that form a cycle, returning it to any arbitrarily selected state.  Quantum heat engines, in contrast, operate by passing quantum matter through a closed series of quantum thermodynamic processes \cite{Quan}.  For instance, Kieu \cite{Kieu1, Kieu2} introduced a class of quantum heat engines which consists of two-energy-eigenstate systems (qubits) undergoing, respectively quantum adiabatic processes and energy exchanges with heat baths at different stages of a cycle - the quantum version of the Otto cycle.  Recently, Zhang {\em et al.} \cite{Zhang} extended Kieu's work by considering the quantum Otto engine with a two-qubit (isotropic) Heisenberg XXX chain as the working substance.  The chain is subject to a constant external magnetic field.  The purpose of their paper is to analyze the effect of quantum entanglement on the efficiency of the quantum Otto engine.  This is an important and intriguing development since it brings together concepts from quantum mechanics and thermodynamics - two seemingly different fundamental areas of physics.

Entanglement is a property associated with the state of a composite quantum system made up of at least two subsystems.  It is a nonlocal correlation between quantum systems that does not exist classically.  Therefore, it is imperative to raise the following questions.  What is the relationship between the net positive work derivable from the subsystems and that from the total system?  What is the role of entanglement in this relationship?  In order to obtain plausible answers to these questions, we need a means to calculate the work derivable from a subsystem.  In this paper, we propose information theoretic answers \cite{Nielsen}.  As a concrete example, we consider the two-qubit XY model \cite{Kamta}.  The Hamiltonian for the two-qubit XY chain in an external magnetic field $B_m$ along the $z$ axis is given by
\begin{equation}
H = \frac{1}{2}(1 + \gamma)J\sigma^1_A \otimes \sigma^1_B + \frac{1}{2}(1 - \gamma)J\sigma^2_A \otimes \sigma^2_B + \frac{1}{2}B_m(\sigma^3_A \otimes \sigma^0_B + \sigma^0_A \otimes \sigma^3_B),
\end{equation}
where $\sigma^0_{\alpha}$ is the identity matrix and $\sigma^i_{\alpha}$ $(i = 1, 2, 3)$ are the Pauli matrices at site $\alpha = A, B$.  The parameter $-1 \leq \gamma \leq 1$ measures the anisotropy of the system.  $(1 + \gamma)J$ and $(1 - \gamma)J$ are real coupling constants for the spin interaction.  The chain is said to be antiferromagnetic for $J > 0$ and ferromagnetic for $J < 0$.  Here, we consider $J > 0$.

To set the stage, we describe the four quantum thermodynamic processes of the quantum Otto cycle.  In the following, we consider as working substance a bipartite quantum system consisting of two subsystems, $A$ and $B$, with Hamiltonian $H = \sum_iE_i|\Psi^i\rangle_{AB}\langle\Psi^i|$.  Here, $E_i$ and $|\Psi^i\rangle_{AB}$ are respectively the eigenvalues and eigenvectors of $H$.  For the two-qubit XY model, we have
\begin{eqnarray}
|\Psi^1\rangle_{AB} & = & \frac{1}{\sqrt{({\cal B} + B_m)^2 + \gamma^2J^2}}
[({\cal B} + B_m)|00\rangle_{AB} + \gamma J|11\rangle_{AB}], \nonumber \\
|\Psi^2\rangle_{AB} & = & \frac{1}{\sqrt{2}}[|01\rangle_{AB} + |10\rangle_{AB}], \nonumber \\
|\Psi^3\rangle_{AB} & = & \frac{1}{\sqrt{2}}[|01\rangle_{AB} - |10\rangle_{AB}], \nonumber \\
|\Psi^4\rangle_{AB} & = & \frac{1}{\sqrt{({\cal B} - B_m)^2 + \gamma^2J^2}}
[({\cal B} - B_m)|00\rangle_{AB} - \gamma J|11\rangle_{AB}],
\end{eqnarray}
with $E_1 = -E_4 = {\cal B}$ and $E_2 = -E_3 = J$, where ${\cal B} \equiv \sqrt{B^2_m + \gamma^2J^2}$.  Furthermore, we assume that the system is allowed to thermalize with the heat baths in processes 2 and 4.  Specifically, process 4 brings the system to its initial quantum state given by the density operator
\begin{equation}\label{state1}
\rho^{(1)}_{AB} = \sum_ip_{i1}|\Psi^{i1}\rangle_{AB}\langle\Psi^{i1}|,
\end{equation}
with $p_{i1} \equiv \exp(-E_{i1}/kT_1)/Z_1$ and $Z_1 \equiv \sum_i \exp(-E_{i1}/kT_1)$.  That is, the system is initially in thermal equilibrium with a heat bath at temperature $T_1$.  For the two-qubit XY model, we have $E_{i1} = E_i$ and $|\Psi^{i1}\rangle_{AB} = |\Psi^i\rangle_{AB}$ with $J = J_1$.
\begin{enumerate}
\item  The system is isolated from the heat bath and undergoes a quantum adiabatic process, with for instance $J$ changing from $J_1$ to $J_2$.  Provided the rate of change is sufficiently slow, $p_{i1}$'s are maintained throughout according to the quantum adiabatic theorem \cite{Messiah}.  At the end of process 1, the system has the probability $p_{i1}$ in the eigenstate $|\Psi^{i2}\rangle_{AB}$.  According to Kieu \cite{Kieu1, Kieu2}, an amount of work is performed by the system, but no heat is transferred during this process.
\item The system is brought into some kind of contact with a heat bath at temperature $T_2 < T_1$.  After the irreversible thermalization process, the quantum state of the system is given by the density operator
\begin{equation}\label{state2}
\rho^{(2)}_{AB} = \sum_ip_{i2}|\Psi^{i2}\rangle_{AB}\langle\Psi^{i2}|,
\end{equation}
where $p_{i2} \equiv \exp(-E_{i2}/kT_2)/Z_2$ and $Z_2 \equiv \sum_i \exp(-E_{i2}/kT_2)$.  Here, for the two-qubit XY model, we have $E_{i2} = E_i$ and $|\Psi^{i2}\rangle_{AB} = |\Psi^i\rangle_{AB}$ with $J = J_2$.  It follows from \cite{Kieu1, Kieu2} that only heat is transferred in this process to yield a change in the occupation probabilities, and the heat transferred is given by
\begin{equation}
Q_2 = \sum_iE_{i2}(p_{i2} - p_{i1}).
\end{equation}
\item The system is removed from the heat bath and undergoes a quantum adiabatic process, with for instance $J$ changing from $J_2$ to $J_1$.  At the end of process 3, the system has the probability $p_{i2}$ in the eigenstate $|\Psi^{i1}\rangle_{AB}$.  An amount of work is performed on the system, but no heat is transferred during process 3.
\item The system is brought into some kind of contact with a heat bath at temperature $T_1$.  After the irreversible thermalization process, the quantum state of the system is returned to the initial one in Eq.(\ref{state1}).  The heat transferred in process 4 is given by
\begin{equation}
Q_4 = \sum_iE_{i1}(p_{i1} - p_{i2}).
\end{equation}
\end{enumerate}

From the first law of thermodynamics, the net work done by the quantum Otto engine is \cite{Kieu1, Kieu2}
\begin{eqnarray}\label{WABo}
W_{AB} & = & Q_2 + Q_4 \nonumber \\
       & = & \sum_i(E_{i1} - E_{i2})(p_{i1} - p_{i2}).
\end{eqnarray}
Substituting $E_{ij} = -kT_j\log(p_{ij}Z_j)$ into Eq.(\ref{WABo}) and setting the Boltzmann constant $k \equiv \log_2e$, we obtain
\begin{equation}\label{WAB}
W_{AB} = (T_1 - T_2)\{S[\rho^{(1)}_{AB}] - S[\rho^{(2)}_{AB}]\} - T_1H[p_{i2}||p_{i1}] - T_2H[p_{i1}||p_{i2}].
\end{equation}
Here, $S[\rho^{(j)}_{AB}]$ is the von Neumann entropy of the quantum state $\rho^{(j)}_{AB}$, and $H[p_{i2}||p_{i1}]$ and $H[p_{i1}||p_{i2}]$ are the relative entropies of $p_{i2}$ to $p_{i1}$ and $p_{i1}$ to $p_{i2}$ respectively.  It follows from the non-negativity of the relative entropy that to derive positive work we not only have to demand $T_1 > T_2$ but also $S[\rho^{(1)}_{AB}] > S[\rho^{(2)}_{AB}]$ such that $H[p_{i2}||p_{i1}]$ and $H[p_{i1}||p_{i2}]$ are not too large.  This is true regardless of whether $AB$ is a single quantum system or one composed of two or more subsystems.  We shall illustrate this after the following remark.

We note that $H[p_{i2}||p_{i1}] = S[\rho^{(2)}_{AB}||\rho^{(1)}_{AB}]$, the quantum relative entropy of $\rho^{(2)}_{AB}$ to $\rho^{(1)}_{AB}$ if they share the same eigenstates.   Similarly, we have $H[p_{i1}||p_{i2}] = S[\rho^{(1)}_{AB}||\rho^{(2)}_{AB}]$.  This is the case for the two-qubit XY model when we let $B_m = \eta J$, with $\eta$ some fixed constant.  We assume this to hold from here on without loss of generality.  It follows immediately from Eq.(\ref{WABo}) that $W_{AB}$ is directly proportional to $(J_1 - J_2)$ for any given $\gamma$ and $\eta$.  Another consequence is that $\rho^{(1)}_{AB}$'s and $\rho^{(2)}_{AB}$'s depend solely on $J_1/T_1$ and $J_2/T_2$ respectively.  It then follows from Eq.(\ref{WAB}) that $W_{AB} = 0$ when $J_2/T_2 = J_1/T_1$.  So, in order to have positive $W_{AB}$, we require $(T_2/T_1)J_1 \equiv J_{\min} < J_2 < J_1$.

It is clear, from Eq.(\ref{WAB}),  that the condition $J_{\min} < J_2$ is one on the quantum states $\rho^{(1)}_{AB}$ and $\rho^{(2)}_{AB}$.  It is thus natural to express this condition in terms of quantities that describe the two-qubit states.  Here, we recall the quantum mutual information between the two subsystems $A$ and $B$, ${\cal I}^{(j)}(A:B) \equiv S[\rho^{(j)}_A] + S[\rho^{(j)}_B] - S[\rho^{(j)}_{AB}]$ \cite{Nielsen}.  This is usually used to measure the total correlations between $A$ and $B$ \cite{Henderson}.  Hence, in order to derive positive $W_{AB}$ we demand that ${\cal I}^{(2)}(A:B) > {\cal I}^{(1)}(A:B) \equiv {\cal I}^{(2)}_{\min}$.  If the states are not separable, we may require that ${\cal C}[\rho^{(2)}_{AB}] > {\cal C}[\rho^{(1)}_{AB}] \equiv {\cal C}^{(2)}_{\min}$, where ${\cal C}[\rho^{(j)}_{AB}]$ is the Wootters concurrence \cite{Wootters} associated with $\rho^{(j)}_{AB}$.  This measures the amount of entanglement or nonlocal quantum correlation between $A$ and $B$.  Given $T_1$ and $J_1$, there is therefore a minimum ${\cal I}^{(2)}_{\min}$ or ${\cal C}^{(2)}_{\min}$ below which the quantum Otto engine does not yield any positive work.

Equation (\ref{WAB}) gives explicitly the dependence of $W_{AB}$ on the bath temperatures $T_1$ and $T_2$, and on the quantum states $\rho^{(1)}_{AB}$ and $\rho^{(2)}_{AB}$.  Consider some $T_2$, $J_{\min} < J_2 < J_1$ and $\kappa T_2$, $\kappa J_2$, which yield identical $\rho^{(2)}_{AB}$.  Here, $\kappa$ is a positive constant such that $\kappa J_2 > J_1$.  Therefore, $W_{AB}$ is positive in the first case but negative in the latter since $\kappa J_2 > J_1$.  This difference is obviously due to the bath temperatures.  Our focus here is on analyzing the dependence of positive $W_{AB}$ on the quantum properties of the states.  Therefore, for given $T_1$ and $J_1$, it suffices to chose an appropriately small $T_2$ such that as $J_2$ is increased from $J_{\min}$ to $J_1$, we have states $\rho^{(2)}_{AB}$, with quantum mutual information going from ${\cal I}^{(2)}_{\min}$ to sufficiently close to the maximum possible of $2$ (or with concurrence going from ${\cal C}^{(2)}_{\min}$ to sufficiently close to the maximum possible of $1$).

\begin{figure}
\includegraphics[scale=0.5]{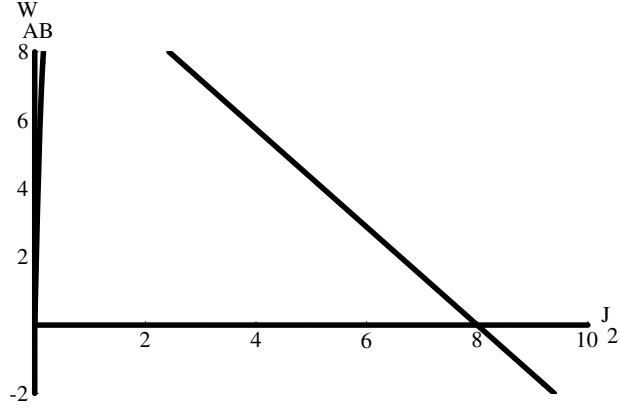}
\caption{Net work done $W_{AB}$ by the quantum Otto engine with a two-qubit XY chain as the working substance plotted vs $J_2$, for $\gamma = 0.4$, $\eta = 0.3$, $T_1 = 1000$, $J_1 = 8$, and $T_2 = 0.1$.}
\end{figure}

\begin{figure}
\includegraphics[scale=0.4]{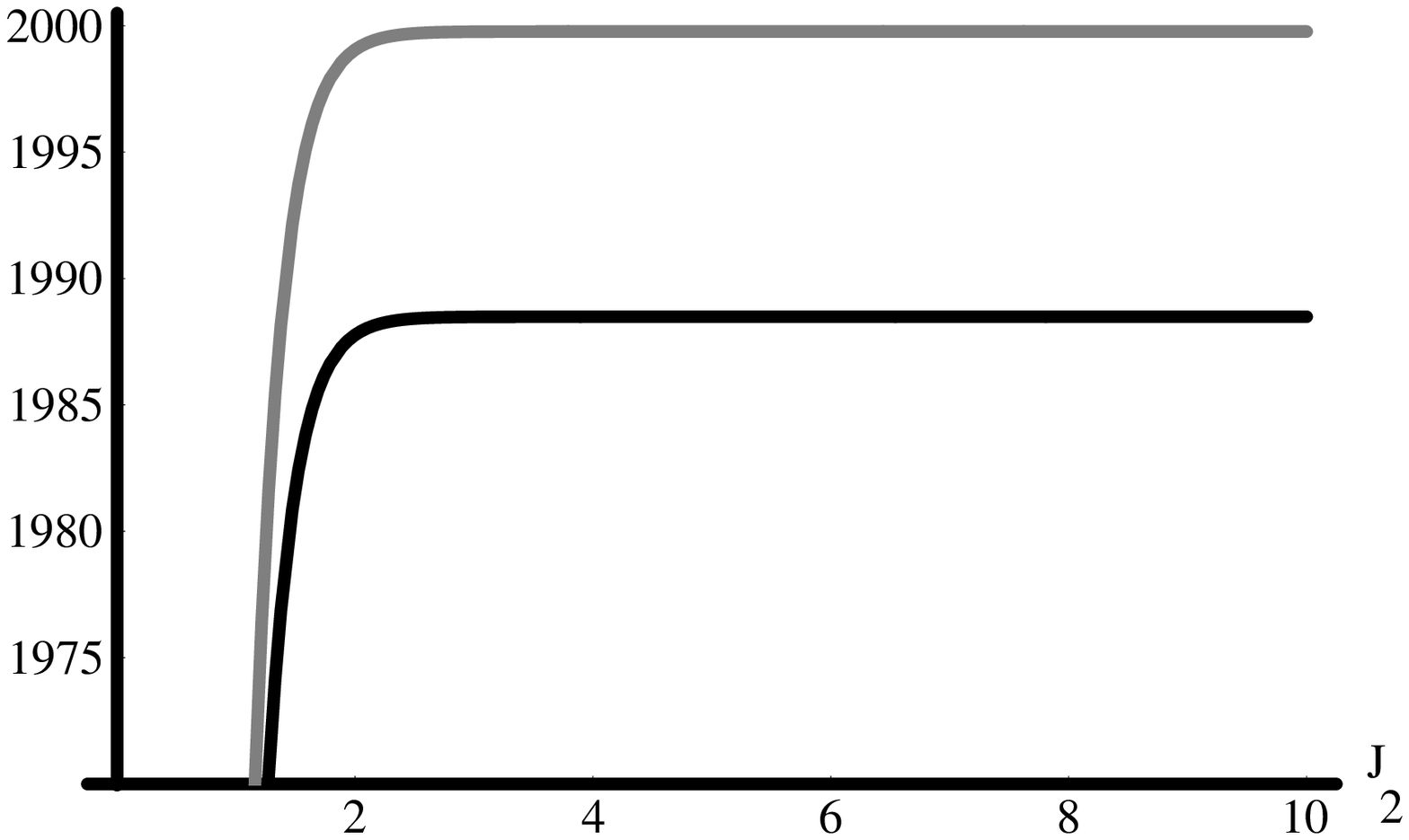}
\includegraphics[scale=0.4]{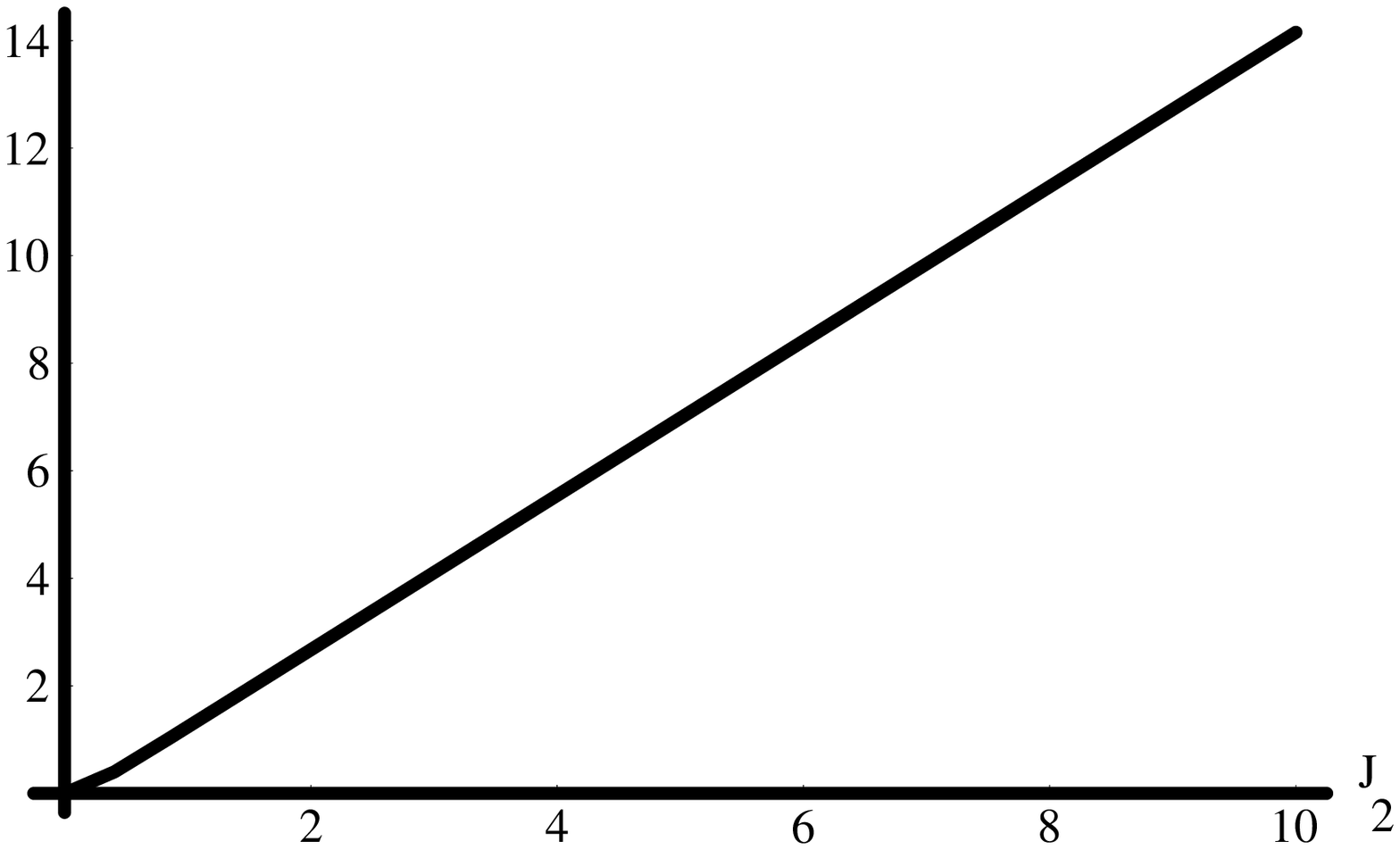}
\caption{The figure on the left hand side shows the behaviour of $(T_1-T_2)\{S[\rho_{AB}^{(1)}]-S[\rho_{AB}^{(2)}]\}$ (gray) and $T_1H[p_{i2}||p_{i1}]$ (black), while that on the right hand side shows the behaviour of $T_2H[p_{i1}||p_{i2}]$.}
\end{figure}

Equation (\ref{WAB}) also provides quantum information theoretic insights into the condition $J_2 < J_1$.  Suppose $\gamma = 0.4$, $\eta = 0.3$, $T_1 = 1000$, $J_1 = 8$, and $T_2 = 0.1$.  Then, $p_{11} \approx p_{21} \approx p_{31} \approx p_{41} \approx 0.25$ and $\rho^{(1)}_{AB}$ approximates the density operator of a maximally mixed state with neither classical nor quantum correlations.  We also note that ${\cal I}^{(2)}(A:B) \approx 2$ and ${\cal C}[\rho^{(2)}_{AB}] \approx 1$ when $J_2 = 8$, satisfying the above sufficient condition.  Figure 1 shows the dependence of $W_{AB}$ on $J_2$.  $W_{AB}$ increases from zero at $J_2 = J_{\min} = 8 \times 10^{-4}$ to a maximum $W_{\max} \approx 10.3695$ at $J_2 = J_{\max} \approx 0.575065$, after which it decreases from $W_{\max}$ to zero at $J_2 = J_1$.  Increase in $J_2$ yields $\rho^{(2)}_{AB}$ with $(T_1 - T_2)\{S[\rho^{(1)}_{AB}] - S[\rho^{(2)}_{AB}]\}$ and $T_1H[p_{i2}||p_{i1}]$ approaching the approximate maxima $1999.81$ and $1988.48$ respectively, but with $T_2H[p_{i1}||p_{i2}]$ that increases monotonically with $J_2$ (see Fig. 2).  At $J_2 = 8$, $T_2H[p_{i1}||p_{i2}]$ exactly equals the difference between the maxima.  And, $W_{AB}$ becomes negative if $J_2$ is increased beyond $J_2 = J_1$.  Intuitively, one would expect to draw more work as the von Neumann entropy of $\rho^{(2)}_{AB}$ differs more from that of $\rho^{(1)}_{AB}$; for instance, when $\rho^{(2)}_{AB}$ becomes increasingly correlated.  This is indeed the case before this difference results in significant increases in the relative entropies enough to cause $W_{AB}$ to decrease to zero.  For $J_{\min} < J_2 < J_{\max}$, $W_{AB}$ increases with increasing quantum mutual information (compare with Fig. 3).  During this increase the correlation changes from purely classical to a mixture of classical and increasingly quantum ones (see Fig. 4).  Beyond $J_2 = J_{\max}$ when ${\cal C}[\rho^{(2)}_{AB}] \approx 0.903444$, $W_{AB}$ begins to decrease.  It is not obvious at this stage what precisely is the role of entanglement.  We shall further elaborate on this after we propose the following definition for the work derivable from the individual subsystem.

\begin{figure}
\includegraphics[scale=0.5]{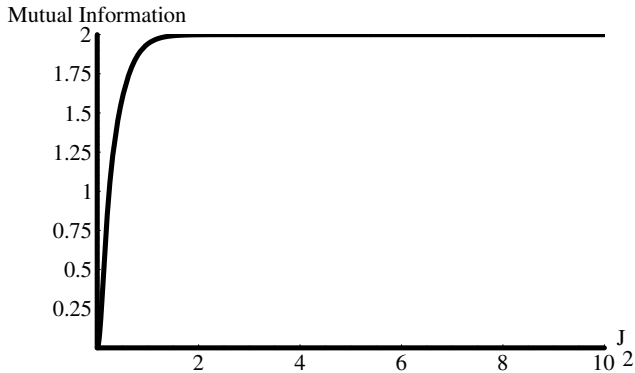}
\caption{Quantum mutual information ${\cal I}[\rho^{(2)}_{AB}]$ in the two-qubit XY chain plotted vs $J_2$, for $\gamma = 0.4$, $\eta = 0.3$, and $T_2 = 0.1$.  At $J_2 = 0$, it is zero since the qubits do not interact.  It then increases with increasing $J_2$ to a maximum of $2\log2$ where the correlation is completely quantum in nature.}
\end{figure}

\begin{figure}
\includegraphics[scale=0.5]{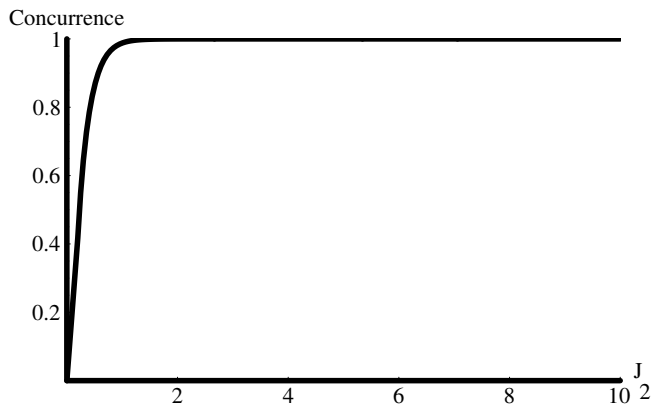}
\caption{The state $\rho^{(2)}_{AB}$ is separable when $J_{\min} < J_2 < 9.31358 \times 10^{-2}$.  Otherwise, it is entangled - with the concurrence approaching one as $J_2$ increases since $p_{32} \rightarrow 1$ and $p_{12} \approx p_{22} \approx p_{42} \rightarrow 0$ with increasing $J_2$.}
\end{figure}

The subsystems $A$ and $B$ are clearly being subjected to exactly the same quantum Otto cycle that the composite system $AB$ has undergone.  The quantum states of $A$ and $B$ that correspond to Eqs.(\ref{state1}) and (\ref{state2}) are also well-defined.  Namely, $\rho^{(j)}_A = {\rm tr}_B\rho^{(j)}_{AB}$ and $\rho^{(j)}_B = {\rm tr}_A\rho^{(j)}_{AB}$.  Hence, inspired by Eq.(\ref{WAB}), we define
\begin{equation}\label{walpha}
w_{\alpha} \equiv (T_1 - T_2)\{S[\rho^{(1)}_{\alpha}] - S[\rho^{(2)}_{\alpha}]\} 
- T_1H[q^{(\alpha)}_{i2}||q^{(\alpha)}_{i1}] - T_2H[q^{(\alpha)}_{i1}||q^{(\alpha)}_{i2}],
\end{equation}
where $\alpha = A$ or $B$, $q^{(A)}_{ij}$'s and $q^{(B)}_{ij}$'s are the eigenvalues of $\rho^{(j)}_A$ and $\rho^{(j)}_B$ respectively.  Now, consider $\rho^{(1)}_{AB}$ approximately maximally mixed and $\rho^{(2)}_{AB}$ sufficiently close to a maximally entangled state, then both $\rho^{(1)}_{\alpha}$ and $\rho^{(2)}_{\alpha}$ will be almost identical to the maximally mixed state.  According to Eq.(\ref{WAB}), the work derivable from a subsystem undergoing the quantum Otto cycle in this case will be extremely close to zero.  For the two-qubit XY model, $\rho^{(j)}_A$ and $\rho^{(j)}_B$ are identical.  Figure 5 shows $w_A$ as a function of $J_2$.  $w_A$ increases from zero at $J_2 = J_{\min}$ to a maximum $w_{\max} \approx 0.231128$ at $J_2 = j_{\max} \approx 0.166289$, when ${\cal C}[\rho^{(2)}_{AB}] \approx 0.316188$.  After $J_2 = j_{\max}$, $w_A$ decreases from $w_{\max}$ to zero at $J_2 = j_{\rm crit} \approx 1.24252$ when ${\cal C}[\rho^{(2)}_{AB}] = {\cal C}^{(2)}_{\rm crit} \approx 0.9964$.  Therefore, Eq.(\ref{walpha}) quantitatively describes the above consideration.  We propose here that $w_A$ ($w_B$) is the work derivable from the subsystem $A$ ($B$).  Therefore, given $T_1$ and $J_1$, there is a critical concurrence ${\cal C}^{(2)}_{\rm crit}$ above which no positive work can be drawn locally from each subsystem.  Increasing $J_2$ beyond $j_{\rm crit}$, $w_A$ becomes negative.  This is because $\rho^{(2)}_A$ becomes more random than $\rho^{(1)}_A$.

\begin{figure}
\includegraphics[scale=0.5]{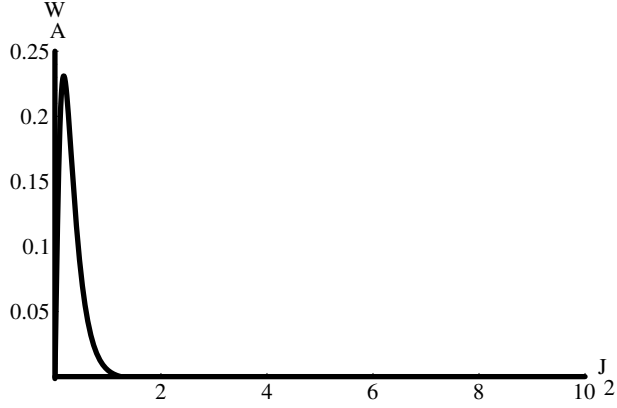}
\caption{Net work done $w_A$ by the quantum Otto engine with one qubit of a two-qubit XY chain as the working substance plotted vs $J_2$, for $\gamma = 0.4$, $\eta = 0.3$, $T_1 = 1000$, $J_1 = 8$, and $T_2 = 0.1$.}
\end{figure}

In conclusion, we have expressed, in Eq.(\ref{WAB}), the net work done $W_{AB}$ by a quantum Otto engine explicitly in terms of the macroscopic bath temperatures ($T_1$ and $T_2$) and information theoretic quantities associated with the microscopic quantum states ($\rho^{(1)}_{AB}$ and $\rho^{(2)}_{AB}$) of the working substance - a bipartite quantum system.  Equation (\ref{WAB}) inspires our proposal of a plausible formula, Eq.(\ref{walpha}), for the work derivable from the subsystems $A$ and $B$.  For the two-qubit XY chain, we show that $\rho^{(2)}_{AB}$ must have quantum mutual information ${\cal I}^{(2)}(A:B)$ or concurrence ${\cal C}[\rho^{(2)}_{AB}]$ greater than that of a given $\rho^{(1)}_{AB}$ to yield positive $W_{AB}$.  Equation (\ref{WAB}) also provides information theoretic insights into how $W_{AB}$ increases and then decreases with increasing ${\cal I}^{(2)}(A:B)$ or ${\cal C}[\rho^{(2)}_{AB}]$.  We show, using Eqs.(\ref{WAB}) and (\ref{walpha}), that there exists a critical concurrence above which no positive work can be derived locally from each subsystem while $W_{AB} > 0$.  This could be another interesting manifestation of quantum nonlocality.


\begin{thebibliography}{99}
\bibitem{Quan} H. T. Quan, Yu-xi Liu, C. P. Sun, and Franco Nori, quant-ph/0611275.
\bibitem{Kieu1} T. D. Kieu, Phys. Rev. Lett. {\bf 93}, 140403 (2004).
\bibitem{Kieu2} T. D. Kieu, Eur. Phys. J. D {\bf 39}, 115 (2006).
\bibitem{Zhang} T. Zhang, W.-T. Liu, P.-X. Chen, and C.-Z. Li, Phys. Rev. A {\bf 75}, 062102 (2007).
\bibitem{Nielsen} M. A. Nielsen and I. L. Chuang, 
{\it Quantum Computation and Quantum Information} (Cambridge University Press, Cambridge 2000).
\bibitem{Kamta} G. L. Kamta and A. F. Starace, Phys. Rev. Lett. {\bf 88}, 107901 (2002).
\bibitem{Messiah} A. Messiah, {\it Quantum Mechanics} (Dover, New York, 1999).
\bibitem{Wootters} W. K. Wootters, Phys. Rev. Lett. {\bf 80}, 2245 (1998).
\bibitem{Henderson} L. Henderson and V. Vedral, J. Phys. A {\bf 34}, 6899 (2001).
\end{thebibliography}
\end{document}